\documentclass[twocolumn]{aastex62}

\newcommand{\Msun}{${\rm M}_{\odot}$}

\newcommand{\lya}{Ly$\alpha$}
\newcommand{\lyman}{Lyman-$\alpha$}
\newcommand{\hi}{\ion{H}{1}}

\newcommand{\ciii}{\ion{C}{3}}
\newcommand{\civ}{\ion{C}{4}}

\newcommand{\siiv}{\ion{Si}{4}}

\newcommand{\ovi}{\ion{O}{6}}

\newcommand{\nref}{N_{\rm ref}}

\defcitealias{peeples18}{Paper~I}

\shorttitle{FOGGIE II: $z = 3$ Circumgalactic Emission}
\shortauthors{Corlies et al.}

\begin{document}

\title{Figuring Out Gas \& Galaxies in Enzo (FOGGIE). II.\\ Emission from the $z=3$ Circumgalactic Medium}

\correspondingauthor{Lauren \ Corlies}
\author[0000-0002-0646-1540]{Lauren Corlies}
\affiliation{Department of Physics \& Astronomy, Johns Hopkins University, 3400 N.\ Charles Street, Baltimore, MD 21218}
\affiliation{Large Synoptic Survey Telescope, Tucson, AZ 85719}
\email{lcorlies@lsst.org}

\author[0000-0003-1455-8788]{Molly S.\ Peeples}
\affiliation{Space Telescope Science Institute, 3700 San Martin Drive, Baltimore, MD, 21218}
\affiliation{Department of Physics \& Astronomy, Johns Hopkins University, 3400 N.\ Charles Street, Baltimore, MD 21218}

\author[0000-0002-7982-412X]{Jason Tumlinson}
\affiliation{Space Telescope Science Institute, 3700 San Martin Drive, Baltimore, MD, 21218}
\affiliation{Department of Physics \& Astronomy, Johns Hopkins University, 3400 N.\ Charles Street, Baltimore, MD 21218}

\author[0000-0002-2786-0348]{Brian W.\ O'Shea}
\affiliation{Department of Computational Mathematics, Science and Engineering, Department of Physics and Astronomy, National Superconducting Cyclotron Laboratory, Michigan State University, East Lansing, MI 48823}

\author[0000-0001-9158-0829]{Nicolas Lehner}
\affiliation{Department of Physics, University of Notre Dame, Notre Dame, IN 46556}

\author[0000-0002-2591-3792]{J.\ Christopher Howk}
\affiliation{Department of Physics, University of Notre Dame, Notre Dame, IN 46556}

\author[0000-0002-7893-1054]{John M.\ O'Meara}
\affiliation{Department of Physics, St.\ Michael's College, Colchester, VT 05439}
\affiliation{W.\ M.\ Keck Observatory, Waimea, HI 96743}

\begin{abstract}
Observing the circumgalactic medium (CGM) in emission provides 3D maps of the spatial and kinematic extent of the gas that fuels galaxies and receives their feedback.  We present mock emission-line maps of highly resolved CGM gas from the FOGGIE project (Figuring Out Gas \& Galaxies in Enzo) and link these maps back to physical and spatial properties of the gas. By increasing the spatial resolution alone, the total luminosity of the line emission increases by an order of magnitude. This increase arises in the abundance of dense small-scale structure resolved when the CGM gas is simulated to $\lesssim 100$ pc scales. Current integral field unit instruments like KCWI and MUSE should be able to detect the brightest knots and filaments of such emission, and from this to infer the bulk kinematics of the CGM gas with respect to the galaxy. We conclude that accounting for small-scale structure well below the level of instrument spatial resolution is necessary to properly interpret such observations in terms of the underlying gas structure driving observable emission.
\end{abstract}


\section{Introduction} \label{sec:intro}
Diffuse gas that is within galactic halos but outside the star-forming disk, referred to as the circumgalactic medium (CGM), is critical to how galaxies evolve \citep*{tumlinson17}. This gas is comprised of metal-poor inflows from the intergalactic medium (IGM), metal-rich outflows from supernova (SN), feedback in the galactic disk, and intermediate metallicity gas that is mixed as gas recycles onto the disk or is stripped from in-falling satellite galaxies.  While these processes are all readily seen in simulations, observing them in emission remains difficult because the high temperatures and low densities of the gas shift most of the emission to ultraviolet wavelengths and low surface brightnesses. Using the FOGGIE (Figuring Out Gas \& Galaxies In Enzo) simulations, we show here that better resolving the CGM in simulations has a profound impact on predictions for the surface brightness and kinematics of observable circumgalactic emission.

At low redshift, CGM absorption measurements have been connected to galaxy properties  \citep[][]{stocke13,tumlinson13,werk14,bordoloi14,liang14,borthakur16,keeney18,berg18}.  However, such samples are inherently limited by the number of UV bright quasars needed to make the absorption measurements. 
At high redshift ($z\gtrsim 2$), the absorption lines probing this gas have shifted into visible wavelengths. Studies of damped \lyman\ absorbers (DLAs; \citealp*{wolfe05,neeleman13,rafelski16}), super Lyman limit systems / sub-DLAs \citep{peroux08,som15,fumagalli16,quiret16}, Lyman limit systems (LLSs; \citealp*{lehner14,fumagalli16,lehner16}), and partial LLSs \citep{lehner16} have long shown large amounts of dense \hi\ and corresponding metals throughout the universe. Yet the redshift that puts these absorption lines within reach also shifts key line diagnostics of the associated galaxies into the infrared and out of the range of detection by current instrumentation. Thus, relating the absorption features to their galactic environment at high-$z$ has remained challenging \citep[though see][]{rudie12a,rudie13,turner14,turner15,turner17}. 

In contrast, observing the CGM directly in \emph{emission} promises to help us understand the spatial and physical distribution of the gas around a single galaxy. Yet emission studies have faced similar challenges when trying to resolve their sources. Recently, two powerful new integral field units (IFUs) on 8--10m class telescopes---the Multi Unit Spectroscopic Explorer (MUSE) on the VLT \citep{bacon10} and the Keck Cosmic Web Imager (KCWI) on Keck \citep{morrissey18}---have provided exciting new tools with which to detect spatially extended \lyman\ emission. Looking to quasars as triggers for bright emission from the gas surrounding them, most at $2 < z < 3$, have measurable \lya\ profiles extending as far as 80 kpc from the galaxy on average \citep{arrigoni18b}, while a handful have detected emission as distant as 200--300 kpc \citep{borisova16,arrigoni18a,cai18}. For galaxies, MUSE has revealed \lya\ around nearly every galaxy it has observed in the high-$z$ universe \citep{wisotzki18} though generally for a smaller median extent of 4--5 kpc \citep{leclercq17}. Though the source of this ionization is still unclear \citep{prescott15}, IFUs probe the dynamics of the gas, showing both inflows \citep{martin14} and outflows \citep{swinbank15} from galaxies.

While \lya\ can tell us much about the CGM, there are advantages to searching for the dimmer emission driven by metal lines. First, because \lya\ is a resonant line, untangling the structure of the emitting gas versus the gas scattering \lya\ photons is challenging and requires modeling of the radiative transfer. In addition, \lya\ necessarily traces the relatively cool, dense gas preferred by \hi. Metal lines, on the other hand, can probe the full range of densities, temperatures, and ionization states expected in the CGM because of the large number of available transitions. Metal lines also trace the gas flows that drive galaxy evolution and set the physical properties of the CGM itself. 

Because metal-line emission is very faint in practice, simulations can help guide our search for detectable targets.  Though the CGM has increasingly been used to place novel constraints on the sub-grid physics recipes in hydrodynamic simulations \citep{hummels13,suresh15,suresh17,ford16}, few theoretical emission predictions have been made for a large number of lines at high-$z$ since \citet{bertone12}. \citeauthor{bertone12} established which lines emit the most brightly in the CGM and highlighted the strong dependency of the emission on both the gas density and temperature in relation to the cooling curves of the emitting ions. \citet{sravan16} explored the variable nature of CGM emission and discussed how detectable emission will be biased towards galaxies having recently experienced large starburst events.  In their work at low-$z$, \citet{bertone10} also demonstrated the relative insensitivity of emission to changes in the simulation's feedback prescriptions because of its strong bias to high densities. \citet{frank12} highlighted the strength of CGM emission relative to IGM emission, indicating that it was a good candidate for direct detection. \citet{corlies16} focused on low-$z$ emission around a single galaxy and found that the brightest emission follows the filament structure of the halo, and determined that simulation resolution indeed limits the ability to draw physical conclusions.  However, while these studies mention the relevance of the predictions for upcoming instrumentation, only \citet{frank12} makes specific instrument-focused predictions for FIREBall \citep{tuttle08} from their simulations. 

In this paper, we analyze the first generation of the FOGGIE simulations, wherein we take a novel approach where the spatial resolution in the CGM of a Milky Way-like galaxy is forced to be as high as the resolution in the galactic disk, an improvement of 8--$32 \times$ better than what is typically found in similar simulations (though see recent work from \citealp{vandevoort19} and \citealp{suresh18}).   
With this new approach to resolving the CGM, we investigate how our predictions of emission from this gas change due to the resolution alone. In particular, we investigate how the observable properties of the gas change and how they can be linked to changes in the physical properties of the gas. While our focus is on $z=3$ to maximize the number of lines observable by current ground-based IFUs while minimizing the effects of surface brightness dimming, these lessons are broadly application to $2 \lesssim z \lesssim 4$ when the galaxy has passed the first stages of star formation but has not finished merging into the final, massive galaxy.

In Section \ref{sec:sims}, we present the simulations and the refinement method that allows us to achieve such high resolution in the outer halo. In Section \ref{sec:predict}, we make predictions for CGM metal-line emission and examine how the results change with resolution. In Section \ref{sec:phys}, we link the changes in observable properties to changes in the physical, ionization state of the gas. In Section \ref{sect:instr}, we make specific predictions for different observing modes of KCWI and MUSE for easy comparison with future observations. Finally, in Section \ref{sec:disc} we discuss the broader context of our results and summarize our conclusions in Section \ref{sec:conc}.

\section{Simulations and Methods} \label{sec:sims}
The  cosmological hydrodynamic simulations we analyze here are the same as presented in \citet[][hereafter \citetalias{peeples18}]{peeples18}; the full details of the simulations and our novel ``forced refinement'' scheme are given there. We briefly review the highlights in Section~\ref{sec:simdetails}; in Section~\ref{sec:calcemis}, we describe how we calculate emissivities from these simulations.

\subsection{Simulation Basics}\label{sec:simdetails}
The FOGGIE simulations were evolved with with Enzo, an Eulerian adaptive mesh refinement (AMR) grid-based hydrodynamic code \citep{bryan14} using a flat \citet{planck13} $\Lambda$CDM cosmology ($1 - \Omega_{\Lambda} = \Omega_{\rm m} = 0.285$, $\Omega_{\rm b} =0.0461$, $h = 0.695$). 
We focus here on a single halo (named ``Tempest'') selected to ultimately have a Milky Way-like mass at $z = 0$ and no major mergers for $z < 1$. The selected halo has $R_{200} = 31$\,kpc and $M_{200} = 4 \times 10^{10}$ \Msun\ at $z = 3$, with dark matter particle mass  $m_{\mathrm{DM}} = 1.39\times10^6 \ \mathrm{M}_{\odot}$.  This halo resides in a cosmological domain with a size of 100 comoving Mpc$/h$. The AMR is allowed to reach a maximum of 11 levels of refinement, corresponding to a finest spatial resolution of 274 comoving pc or a physical resolution of 68\,pc at $z = 3$.

The simulations include metallicity-dependent cooling and a metagalactic UV background \citep{haardt12} using the Grackle chemistry and cooling library \citep{smith17}. All metals are tracked as a single combined field; thus, particular elemental abundances throughout the paper are calculated assuming solar abundances. We use a \citet{cen06} thermal supernova feedback model, forming stars in gas exceeding a comoving number density of $\simeq 0.1$\,cm$^{-3}$ with a minimum star particle mass of $2 \times 10^4$\,\Msun.  The effects of Type Ia SNe are not included. 

The general aim of AMR simulations is to place refinement in areas that are the most physically interesting. Typically with these types of cosmological zoom-in simulations, the additional refinement is triggered primarily by increases in density, with the goal of best refining the dense, star-forming disk of the galaxy of interest. For each level of refinement, the cell size decreases by a factor of two such that
\begin{equation}
    \mathrm{Cell \ Size} = \frac{\mathrm{Box \ Size}}{\mathrm{Root \ Grid \ Cells}} \times 2^{-\nref},
\end{equation}
where $\nref$ is the level of refinement; our root grid is $256^3$. In our standard AMR simulations, the CGM typically reaches a refinement level of 6--8 while the ISM reaches $\nref=11$. This corresponds to 2.2--0.55 kpc resolution in the CGM at $z=3$. However, as discussed in \citetalias{peeples18}, there are many processes relevant to circumgalactic physics with potentially smaller spatial, the cooling length being the most notable.

This first generation of FOGGIE simulations takes a different approach and  targets cells for refinement based on their spatial location alone.
This ``forced refinement'' scheme follows the targeted galaxy with a cubic box that tracks it through the domain. 
To implement forced refinement, we first run a ``standard'' AMR simulation as described above, writing out snapshots in 20\,Myr increments. The main halo is identified and the coordinates of a 200\,kpc comoving box centered on the galaxy are recorded for each snapshot. 
The simulation is then restarted at $z=4$ with the volume enclosed by this box refined to a minimum refinement level; for our default $\nref=10$ run (the ``high-resolution'' simulation in \citetalias{peeples18}), this corresponds to a fixed resolution of 380  h$^{-1}$ comoving parsec.
We have additionally evolved an $\nref=11$ simulation (190 h$^{-1}$ comoving pc) to $z=2.5$
with a cell size of 380 h$^{-1}$ comoving pc ($\nref=10$) or 190 h$^{-1}$ comoving pc ($\nref=11$).  
 The location of the box is updated every 20 Myr.
At $z=3$, the two highly refined runs have physical spatial resolutions of 137\,pc ($\nref=10$) and 68\,pc ($\nref=11$) respectively. Throughout the rest of this paper, we will reference the normal AMR run as ``standard'' while the two highly refined runs will be referred to by this physical size of the refined CGM cells.

\subsection{Calculating Emissivities}\label{sec:calcemis}
For the densities and temperatures typical of the CGM, the gas cools primarily through collisional excitation followed by radiative decay, leading to a $n^2$ dependence of line emission. For a given line, the brightest emission will therefore come from gas with temperatures that correspond to the peak of that line's cooling curve. \citet{bertone13} shows examples of the cooling curves that dominate cooling of the diffuse universe. 

To calculate the emissivity in each cell, the simulation is post-processed using the photoionization code {\sc cloudy} \citep[version 10.0;][]{ferland98}.  For each cell, the emissivity is calculated using {\sc cloudy} tables parameterized by hydrogen number density ($n_{\mathrm{H}}$), temperature ($T$), and redshift. The metal line emissivity is then scaled linearly by the metallicity of each cell. 

First, we constructed {\sc cloudy} look-up tables of emissivity as a function of temperature ($10^3 < T < 10^8 \mathrm{K}, \ \Delta \log T=0.1$) and hydrogen number density ($10^{-6} < n_{\mathrm{H}} < 10^2$~cm$^{-3}$, $\Delta \log n_{\mathrm{H}}=0.5$). The calculation assumes solar metallicity and abundances.  The grid is then linearly interpolated for every cell to the correct temperature and $n_{\mathrm{H}}$.
Finally, {\sc cloudy} also assumes that the gas is in ionization equilibrium, accounting for both photoionization and collisional ionization. For consistency with \citet{corlies16}, we use the 2005 updated version of \citet{haardt01} as our extragalatic ultraviolet background throughout. 

\section{Predicted Emission Properties} \label{sec:predict}
In this section we make predictions for the distribution of metal-line emission at $z=3$ and demonstrate the role CGM resolution plays on the probability of its detection. We present surface brightness maps for H$\alpha$, \siiv, \ciii, \civ, and \ovi\ in Section~\ref{sect:sb_maps}, the role of angular resolution in Section~\ref{sect:angres}, radial profiles and covering fractions in Section~\ref{sect:covering}, and the kinematic properties in Section~\ref{sect:kinematic}.

\subsection{Surface Brightness Maps}\label{sect:sb_maps}

\begin{figure*}
\centering
\includegraphics[width=0.72\textwidth]{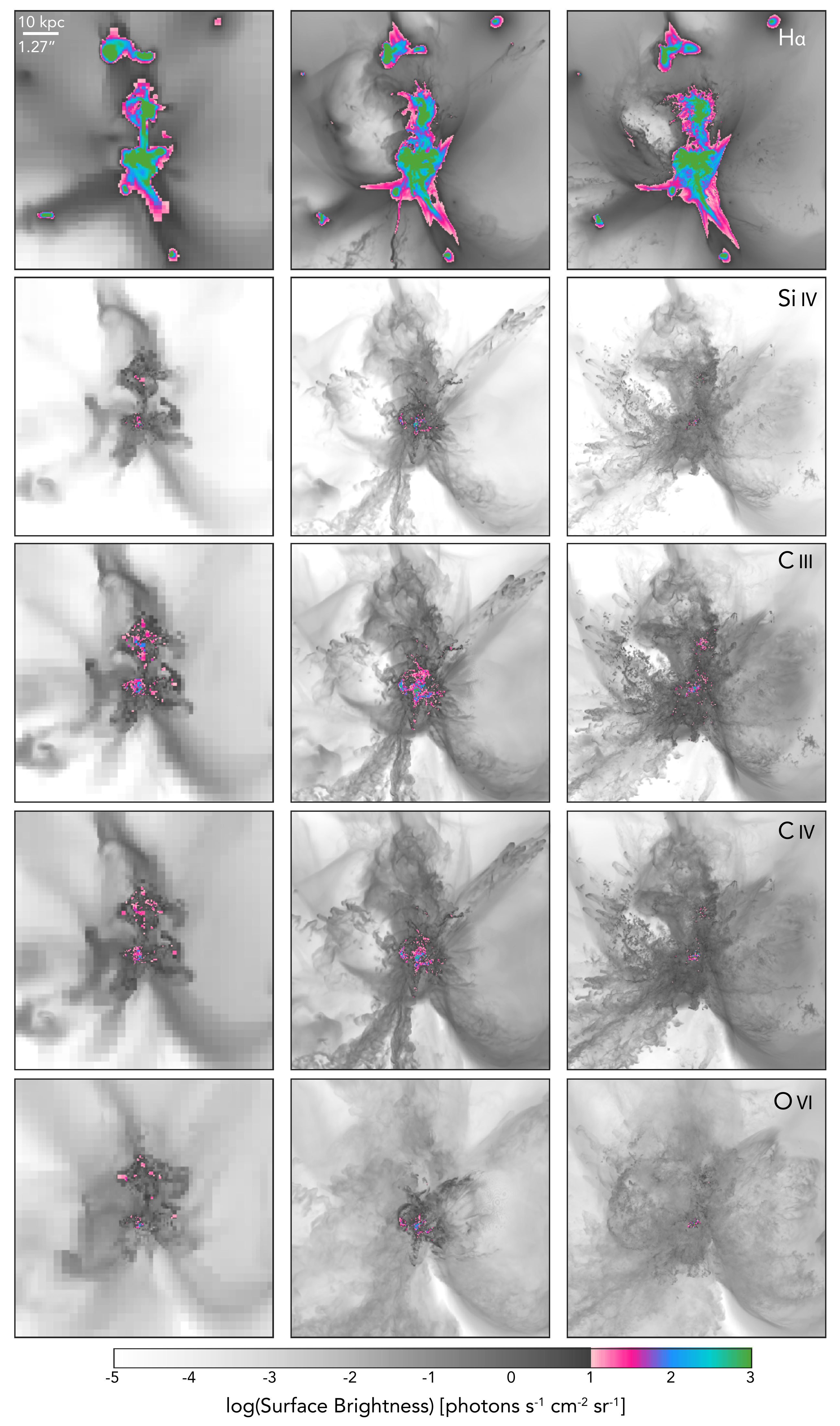}
\caption{Surface brightness maps at $z=3$ of five different emission lines (H$\alpha$, \siiv, \ciii, \civ, and \ovi) for the standard AMR simulation, the 137\,pc simulation, and the 68\,pc simulation. The colors correspond roughly to detection probability with gray being non-detectable and colors related to different levels of likelihood as described in Section \ref{sect:sb_maps}. The pixel size of the standard simulation is 137\,pc and matches the CGM resolution in the two highly refined cases. Denser structures are clearly visible in the more highly refined simulations but most structures will remain beyond the detection limits of current and upcoming instrumentation.  \label{emis_map.fig}}
\end{figure*}

Figure \ref{emis_map.fig} shows surface brightness (SB) maps of the entire 200 comoving kpc high refinement region at $z=3$ for our standard AMR simulation (left), the 137\,pc simulation (middle), and the 68\,pc simulation (right) for H$\alpha$ and a number of metal lines. Because the standard run has varying cell sizes due to the AMR, we choose to force the pixel size to match the 137\,pc simulation for easy comparison. The two highly refined simulations have pixel sizes matching their stated CGM resolution.  The SB dimming of an object at this redshift is accounted for in all images throughout the paper. This colormap will be used throughout the paper and corresponds roughly to the probability of detection with current and upcoming instrumentation.  Green corresponds to pixels that should always be detected (log$_{10}$(SB) $\geq 3$ photons s$^{-1}$ cm$^{-2}$ sr$^{-1}$), blue to pixels that will probably be detected ($2 \leq $ log$_{10}$(SB) $< 3$ photons s$^{-1}$ cm$^{-2}$ sr$^{-1}$), and pink to pixels that are formally possible to detect but push the limits of all instruments ($1 \leq $ log$_{10}$(SB) $< 2$ photons s$^{-1}$ cm$^{-2}$ sr$^{-1}$).  Gray are pixels that will not be detected in the near future (log$_{10}$(SB) $< 1$ photons s$^{-1}$ cm$^{-2}$ sr$^{-1}$).  Detailed matches to two current instruments, KCWI and MUSE, are discussed in Section \ref{sect:instr}.

\begin{figure*}
\centering
\includegraphics[width=0.72\textwidth]{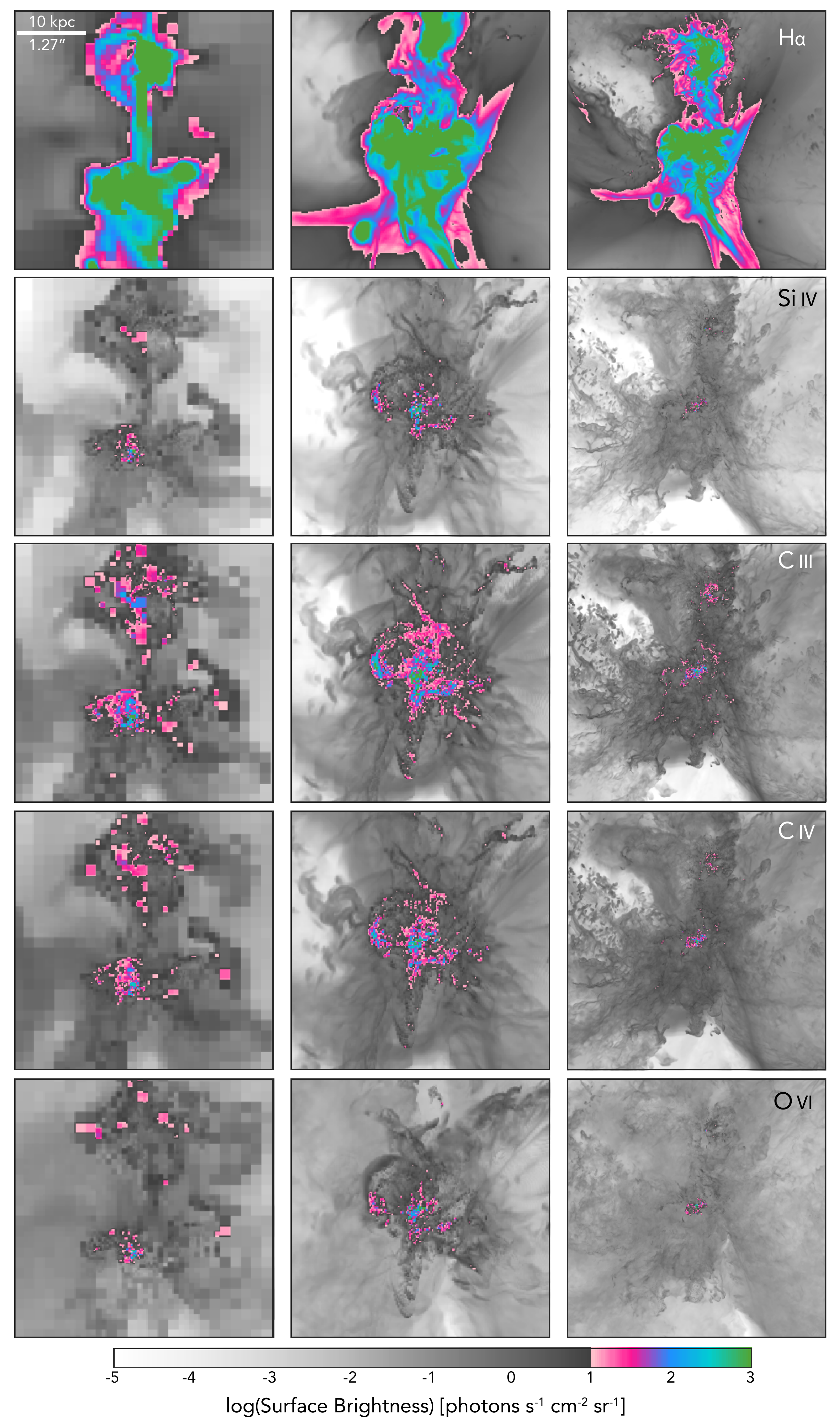}
\caption{Same $z=3$ surface brightness maps as Figure \ref{emis_map.fig} but now zoomed in so only an area of $40\times 40$\,kpc (5\arcsec $\times 5$\arcsec) is shown. The bright, observable emission is confined to within roughly 20 kpc of the galaxy. More disjointed areas can have higher surface brightnesses in the higher resolution simulations where regions are allowed to collapse to higher densities.   \label{emis_map_zoom.fig}}
\end{figure*}

Table \ref{tab:lumin} gives the total luminosity of each line in the 200 kpc comoving refinement region for each of the simulations. While the distribution of the observable emission is not greatly affected by the resolution, the total luminosity emitted in each line does change substantially with resolution: the total luminosity of each emission line we consider {\em increases by about an order of magnitude} owing to improving the circumgalactic resolution alone.
The luminosities from the 68\,pc simulation are much closer to the 137\,pc simulation than either are to the standard-resolution simulation, suggesting that the 137\,pc simulation is nearly converged with respect to this diagnostic.
Improved spatial resolution allows regions in the CGM to collapse to higher density, leading to more efficient cooling radiation and larger luminosities overall. We discuss this effect in more detail in Section \ref{sec:rad_profs}.

\begin{table}
\centering
\begin{tabular}{ |c|c|c|c|c| } 
 \hline
\ Line & Wavelength & Standard & 137\,pc & 68\,pc \\ 
\hline
H$\alpha$ & 6563\,\AA & 8.9e42 & 1.3e43 & 1.2e43 \\
\siiv & 1394\,\AA & 1.6e40 & 4.7e41 & 7.2e42 \\ 
\ciii & 977\,\AA & 1.2e41 & 3.5e42 & 4.5e43 \\ 
\civ  & 1548\,\AA & 8.1e39 & 5.6e41 &  3.2e41 \\
\ovi  & 1032\,\AA & 5.5e39 & 1.4e40 & 2.1e41 \\
\hline
\end{tabular} 
\caption{Total luminosity of a given line within the refinement box for each simulation in units of ergs\,s$^{-1}$. The standard simulation under predicts the luminosity in each line by roughly an order of magnitude compared to the highly refined simulations.} \label{tab:lumin}
\end{table}

In general, lines whose cooling curves peak at slightly lower temperatures, like \ciii,\ tend to be the brightest at this redshift because it is at these temperatures where the bulk of the dense gas throughout the halo is found. \ovi, on the other hand, is particularly weak because there is little dense gas at higher temperatures, resulting in little detectable emission. The physical causes of the emission are addressed further in Section \ref{sec:phys}.

Adding resolution to the CGM clearly reveals the filaments feeding the galaxy and the structure within them that is artificially smoothed by the poor resolution in the standard run (left panels). Other small-scale structure is created by SN-driven outflows and by gas stripped from inflowing satellites. If we want to examine the small scale structure in emission, these highly refined simulations are needed.

However, despite these significant morphological differences between the runs, most of this increased small-scale structure around this relatively small galaxy  is undetectable, as exhibited by the color map.  Almost all of the detectable gas remains within 20 physical kpc, regardless of the CGM resolution.

\begin{figure*}[ht]
\centering
\includegraphics[width=\textwidth]{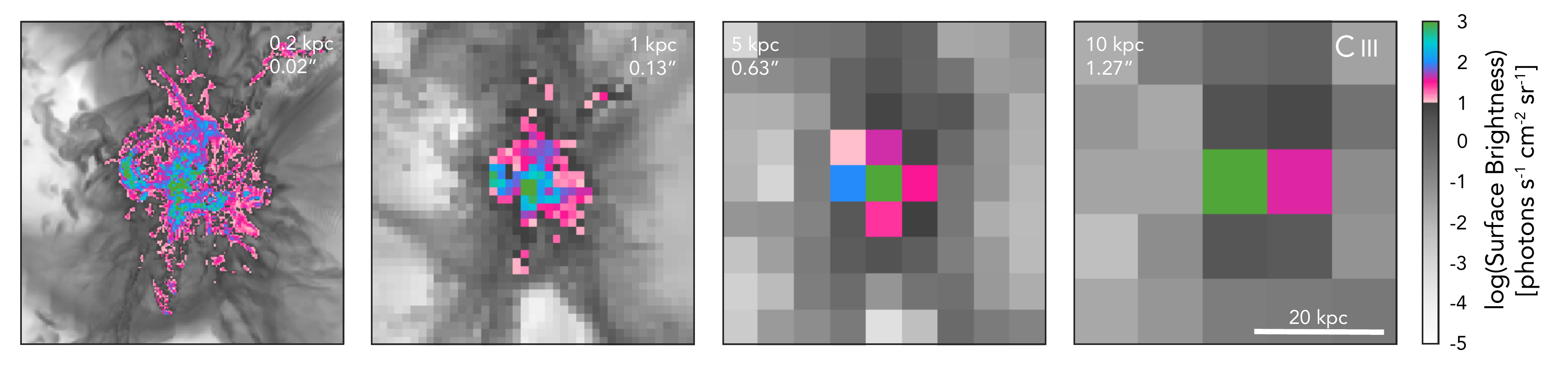}
\caption{Emission map of \ciii\ at $z=3$ for the 137\,pc resolution simulation shown with different angular resolutions, starting with the resolution of the simulated CGM on the left and with pixels getting bigger moving right. For the coarsest angular resolution, a detection can potentially be made. At higher angular resolution, however, not only can the CGM be detected but its underlying structure and the processes shaping it can be probed.  \label{emis_ang_res.fig}}
\end{figure*}

The one large outlier is H$\alpha$.  Because it is independent of metallicity, the line is extremely bright even at $z=3$, tracing the cosmic filaments. However, at $z=3$, H$\alpha$ has shifted to an observed wavelength of 2.6$\micron$, well outside the bandpasses of the ground-based IFUs discussed in this paper. This does fall at a wavelength observable by NIRSpec on the {\em James Webb Space Telescope}; more detailed {\em JWST} predictions will be the focus of future FOGGIE simulations. 

To better highlight the detectable regions, Figure \ref{emis_map_zoom.fig} shows a zoomed in view of the galaxy that is 40 physical kpc across (or 5\arcsec $\times 5$\arcsec\ at $z=3$).  Much of the clearly observable emission is coming from the central part of the galaxy and thus the interstellar medium as opposed to the CGM. Yet, it is also obvious that the higher spatial resolution leads to the formation of small, dense regions that are detectable to larger radii in the 68 pc and 137 pc simulations. Thus, the emission can be clumpy on small scales that would not be predicted if not for this enhanced simulation resolution. The 68 pc and 137 pc simulations show that we can possibly expect to detect emission from the CGM at larger radii from the main galaxy. These will enable us to definitively say that the emission is from the CGM and not from the galaxy itself. 

It is worth noting that the standard simulation displayed here is somewhat unrepresentative. At this particular redshift, the main halo is actually in the process of merging with another dense galaxy which appears as two distinct regions of blue/green pixels in the images. Thus, the extent of emission from the center of either galaxy in these frames is smaller than that from the main galaxy in either highly refined simulation. The orientation of this satellite in the refined simulations is different such that it does not produce as much observable emission.

\subsection{Angular Resolution}\label{sect:angres}

In addition to the surface brightness limits, the angular resolution of an observation can have a large effect on the conclusions that can be drawn about the CGM. Figure \ref{emis_ang_res.fig} takes the \ciii\ emission from the 137\,pc simulation shown in Figure \ref{emis_map_zoom.fig} as the leftmost panel. The angular resolution of the image is then degraded as the panels move towards the right. The labels show physical size and angular size at $z=3$ for each panel. 

For the coarsest resolution shown, the galaxy can barely be detected. At 5\,kpc resolution, both the galaxy and the CGM are likely to be detected but it is difficult to glean any information about the true shape and physical distribution of CGM properties. Instead, the resolution of 1\,kpc (0.13\arcsec) is necessary to discern both the CGM's spatial distribution and also to say anything definitive about the processes that are shaping the CGM. At this resolution, one can see that the gas is not spherically symmetric and that it is clumping on scales of at least the size of the pixels.  This resolution is fine, but is not impossible to achieve with current instrumentation and highlights the need to prioritize high angular resolution in future instrumentation \citep[e.g.,][]{luvoir18}.

\subsection{Surface Brightness Profiles and Covering Fractions}\label{sect:covering}
The emission maps of Figures \ref{emis_map.fig} and \ref{emis_map_zoom.fig} show by eye the differences in the extent and scale of emission in the CGM and how it depends on the simulation resolution. In this section, we quantify these differences with a focus on the observational implications by looking at the radial profile and covering fractions of the surface brightness. 

\begin{figure*}
\centering
\includegraphics[width=0.85\textwidth]{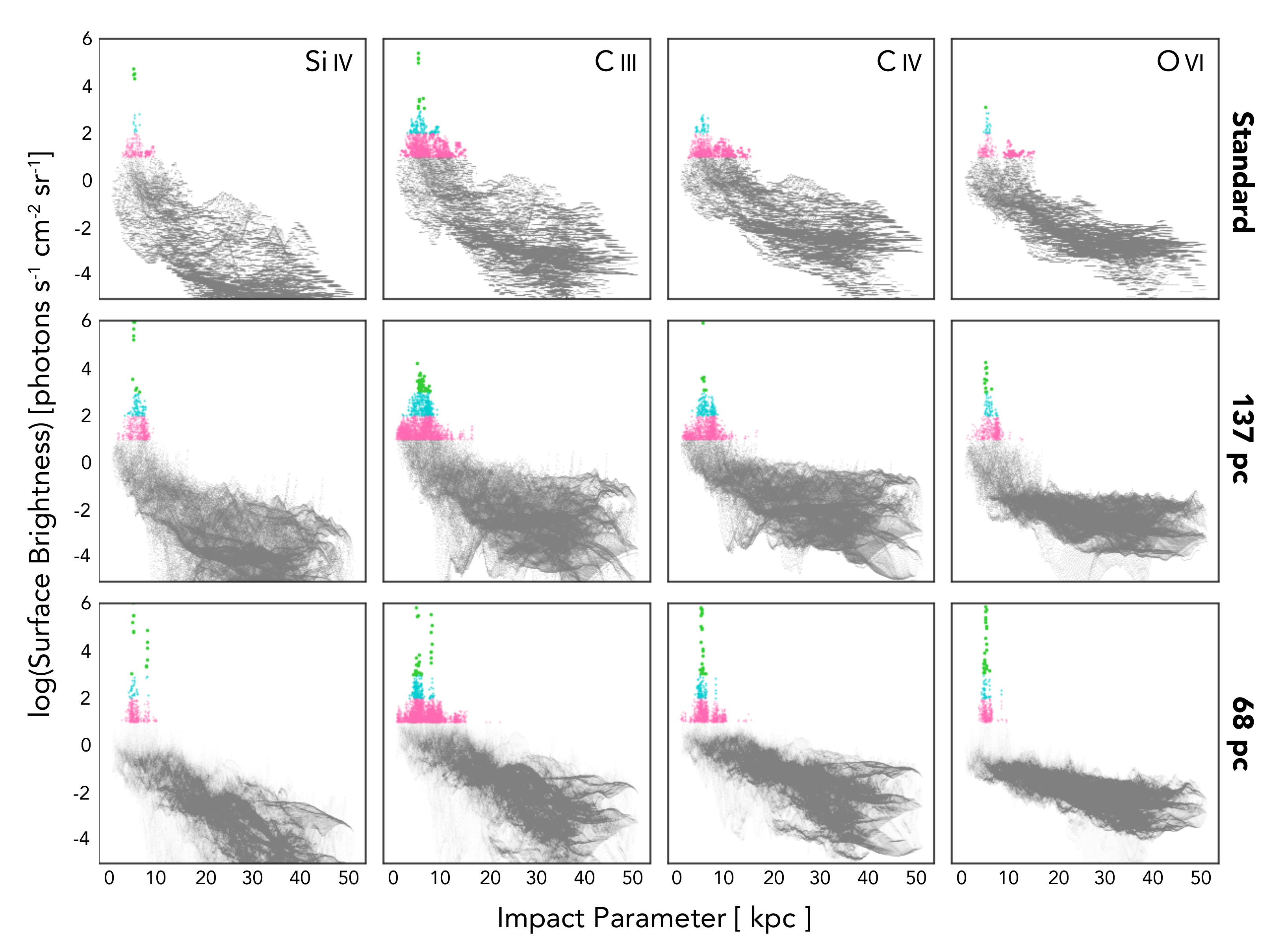}
\caption{Radial profiles of the surface brightness maps shown in Figure \ref{emis_map.fig} for four emission lines and all 3 simulations. The colors correspond to the color maps of Figures \ref{emis_map.fig}--\ref{emis_ang_res.fig}: green -- detectable; blue/pink -- possible to detect; gray -- beyond current instruments. The detectable emission is found within 20 kpc for all the simulations. For non-detectable emission, structures within the CGM gas are much better traced in the simulations with better spatial resolution. \label{sb_prof.fig}} 
\end{figure*}

Figure \ref{sb_prof.fig} takes every pixel shown in the emission maps of Figure \ref{emis_map.fig} and plots the radial profile of the surface brightness for four emission lines for the given projection axis. The colors here are generally matched to the colorbar of Figure \ref{emis_map.fig}. Radial profiles averaging over the three primary simulation axes tend not to show much variation so this single axis is illustrative \citep{corlies16}.  The radial profiles confirm that easily detectable emission is confined to the central parts of the galaxy. However, the potentially detectable (blue and pink) pixels can be found as far as 20\,kpc from the center of the galaxy, allowing for confirmable detection of CGM emission. While most pixels remain undetectable, the radial profiles also highlight how the low resolution in the standard run does not fully sample such low surface brightness structures in the outer CGM. 

Figure \ref{sb_prof.fig}'s emission-focused radial profiles can be compared to the absorption-focused radial profiles in Figure~7 of \citetalias{peeples18}. The emission seems to follow the \hi\ column density the most closely with the brightest emission and the largest \hi\ column densities being found within 20\,kpc of the galaxy. However, the steepness of the SB profiles does not change as strongly with CGM resolution like it did for the column density profiles. This is in part because we have chosen to highlight the detectable emission so the plot spans almost 12 orders of magnitude on the $y$-axis. Looking instead at the undetectable pixels, more pixels exist at a larger spread of values so the radial profile is flatter for these larger radii. However, this similarity to the \hi\ suggests that the main reason for these similar SB profiles is the strong dependence of the emissivity on the gas density whereas the number of pixels traces the volume-filling, diffuse gas.

\begin{figure}
\centering
\includegraphics[width=0.48\textwidth]{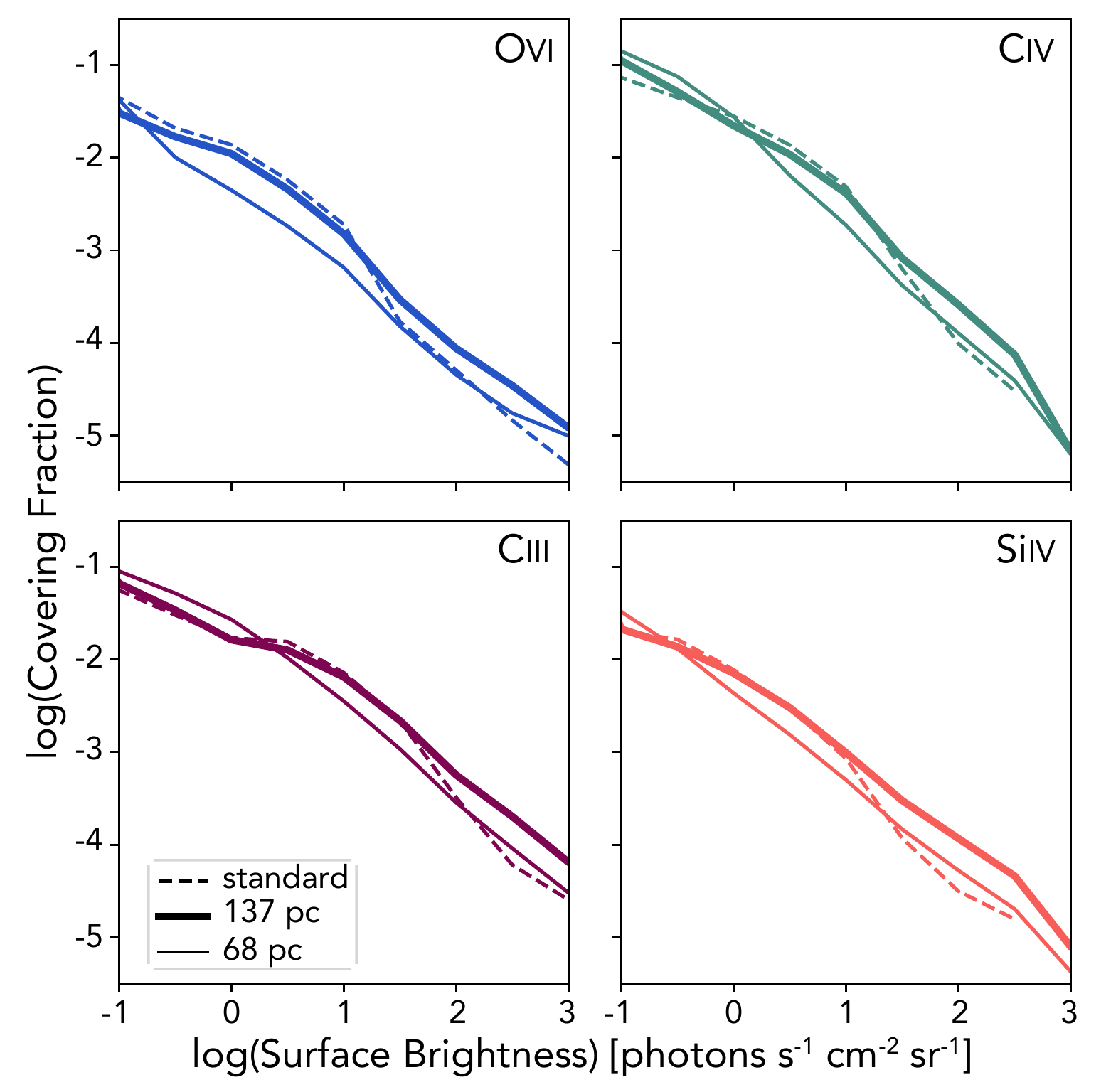}
\caption{Number of pixels above a given surface brightness limit for four emission lines and all three simulations, averaged over the three primary projection axes of the simulation boxes. Fewer than $1\%$ of pixels are observable and the fraction does not vary greatly with the simulation resolution.}
\label{cov_frac.fig}
\end{figure}

We further quantify the observability of the emission by considering covering fractions of varying SB levels. Figure \ref{cov_frac.fig} shows the fraction of pixels above a given surface brightness level for four emission lines for each of the simulations. The covering fraction is then averaged over all three axes of the simulation box to reduce the influence of any preferential viewing angles. 

\subsection{Tracing Kinematic Properties}\label{sect:kinematic}
\begin{figure*}[!ht]
\centering
\includegraphics[width=0.7\textwidth]{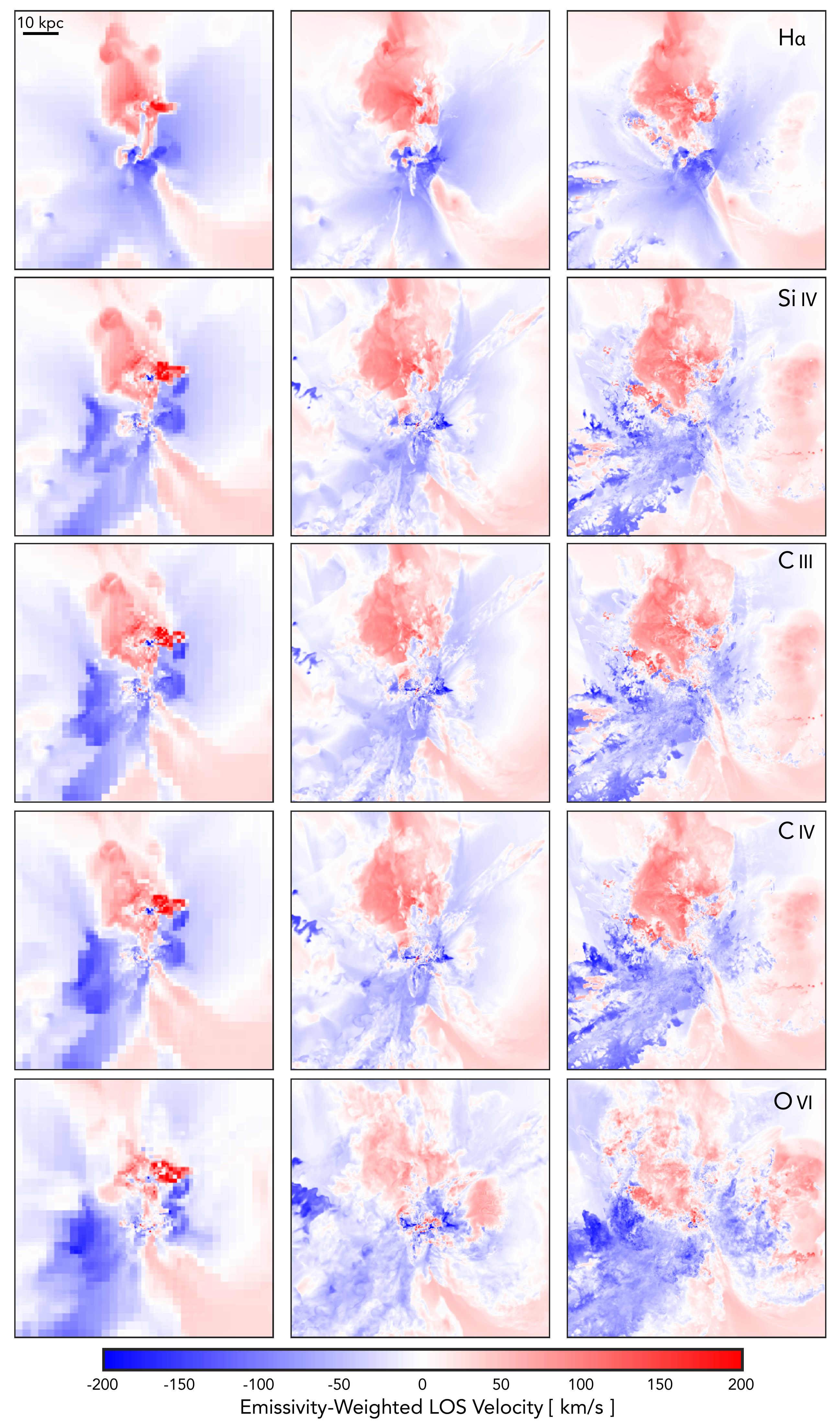}
\caption{Maps of the emissivity-weighted LOS velocity after the bulk velocity of the refinement region has been subtracted. Direct comparisons between simulations is difficult because the orientation of the galaxy changes to match the emission maps shown in Figure \ref{emis_map.fig}. Increasing the resolution increases variations in the kinematics amongst the different emission lines and reveals complex kinematic structures on the smallest scales. \label{velmap.fig}}
\end{figure*}

In general, fewer than $1\%$ of the pixels are detectable for any ion at the highest resolution of each simulation (and binned at 137\,pc for the standard run).  Above 1\,photon\,s$^{-1}$\,cm$^{-2}$\,sr$^{-1}$, the 137\,pc simulation does have a higher covering fraction than the standard simulation. Denser peaks are allowed to form because of the higher spatial resolution, which leads to brighter emission. On the other hand, the 68\,pc simulation has the lowest covering fraction because the bright emission is confined to smaller physical regions which leaves more of the pixels at lower surface brightness. Since each pixel is smaller, the overall number of observable pixels does not decrease (see Figure \ref{sb_prof.fig}) just the percentage of the total. 

A unique strength of using IFUs is that for every pixel, a spectrum is generated, providing kinematic information that can inform our understanding of the gas origins. To begin to estimate such properties from the simulation, we calculate the bulk velocity of the entire refinement region and subtract it from the cells within the box to provide a meaningful frame of reference for the velocities.  Figure \ref{velmap.fig} shows the emissivity-weighted line of sight (LOS) velocity at $z=3$ for each simulation; the projection axis is the same as for the emission maps shown in Figure \ref{emis_map.fig}. 
We caution against directly comparing the simulations because the orientation of the galaxy relative to the projection axis is somewhat different in each simulation. Nevertheless, some general trends can still be identified. 

In the standard simulation, there is not much variation in the velocity structure amongst the different emission lines. In contrast, in the highly refined simulations, while the bulk velocity flows remain similar, more small-scale velocity fluctuations are seen as the ionization energy of the line increases. H$\alpha$ and the other low ions are tracing dense gas which is dominated by coherent filaments at these high redshifts. The higher ions, like \ovi, trace the volume-filling gas which has more peculiar motions from outflows. 

These maps demonstrate how the high resolution in the CGM changes the kinematics which in turn will affect the predicted emission line profiles, akin to the ways we showed how simulated velocity discretization affects absorption line profiles \citepalias{peeples18}. Thus, this resolution is crucial for using simulations to inform the interpretations of future observations of circumgalactic gas kinematics in emission.

\section{Connecting Emission to Physical Conditions}\label{sec:phys}
Ultimately, the goal of observing the CGM in emission is to 
understand the physical properties---the density, temperature, and metallicity---of the gas. In this section, we link the changes in emission properties to changes in the physical properties of the gas. 

\subsection{CGM Physical Properties and Resolution}\label{sec:rad_profs}

\begin{figure*}[!ht]
\centering
\includegraphics[width=0.98\textwidth]{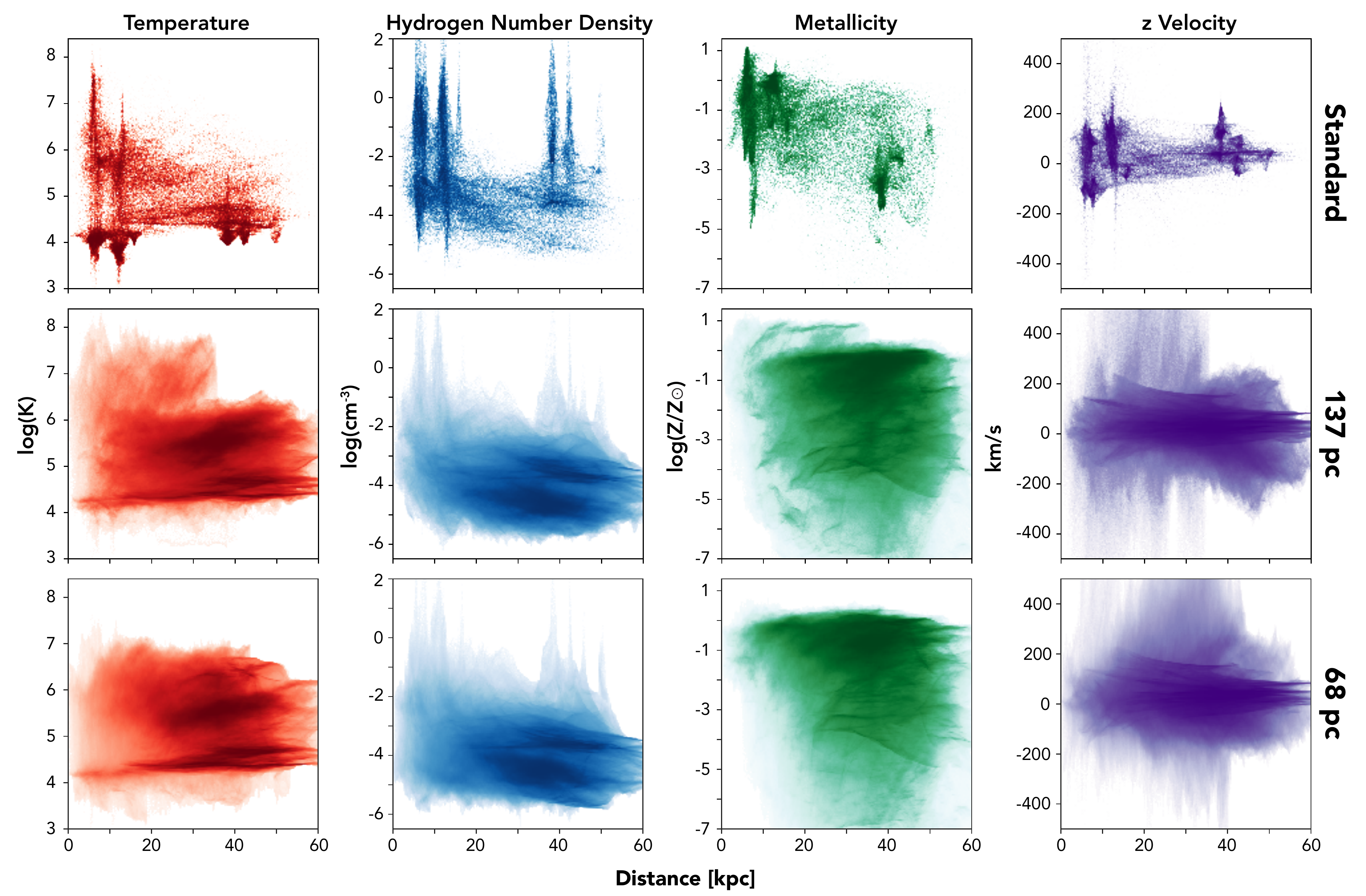}
\caption{Radial profiles at $z=3$ of the temperature, density, metallicity, and z-velocity for the three simulations discussed. Each cell in the volume is depicted such that fewer points corresponds to fewer number of cells in that simulation. While changing the resolution of the CGM does not affect the bulk, average properties of the gas, the spread in all of these physical quantities has changed dramatically at all radii. \label{rad_prof.fig}}
\end{figure*}

Figure \ref{rad_prof.fig} shows the radial profiles of temperature, hydrogen number density, metallicity, and a 1D-velocity for the three simulations presented throughout the paper. In general, the {\em average} physical properties of the gas are unchanged, which is not surprising since all that varies between these simulations is the numerical resolution. However, we do see that the \emph{breadth} of all of these quantities has increased. In the highly resolved CGM, gas can exist at low and high density, temperatures, metallicities, and velocities at all radii. That is, the gas is more multiphase at all radii in this halo at $z=3$ when the CGM is more highly resolved. 

A broader distribution of gas densities means there is more high density gas, which in turn translates directly to the higher total emission noted in Section \ref{sect:sb_maps} and shown in Table \ref{tab:lumin}. Because the emission is predominantly produced by gas cooling through collisional excitation of these lines, the $n^2$ nature of this process means the strength of emission depends strongly on the density. Even though the bulk of the gas remains undetectable, the brightness of the source generically increases because of this more highly resolved, dense gas. 

Similarly, when the gas is more artificially mixed it settles at a single temperature ($\sim 10^{5.5}$\,K in the standard run, for example). This results in the gas cooling more strongly through certain lines (\civ,\ \ovi) at the detriment of others (\siiv,\ \ciii). Instead, the increased resolution allowing gas to be more distributed in temperature means that more gas can also exist at the peak of the cooling curve of a larger number of metal lines. 

Finally, the emissivity of the gas is also regulated by its metallicity. Just as the temperature changes when the gas is artificially mixed, so too does the metallicity. This can help explain why gas is not uniformly brighter in the high resolution simulations with denser gas. If the denser gas also now has lower metallicity, then the emission will not become as bright as gas at the same density but with higher metallicity from the artificial mixing. 

In short, the combination of larger spreads in density, temperature, and metallicity result in more overall emission and in a different spatial distribution of such emission. The complicated interplay of these properties is why emission can be such a useful tool for diagnosing the CGM.

\subsection{Examining the Ionization Process Driving Emission}
\begin{figure}
\centering
\includegraphics[width=0.5\textwidth]{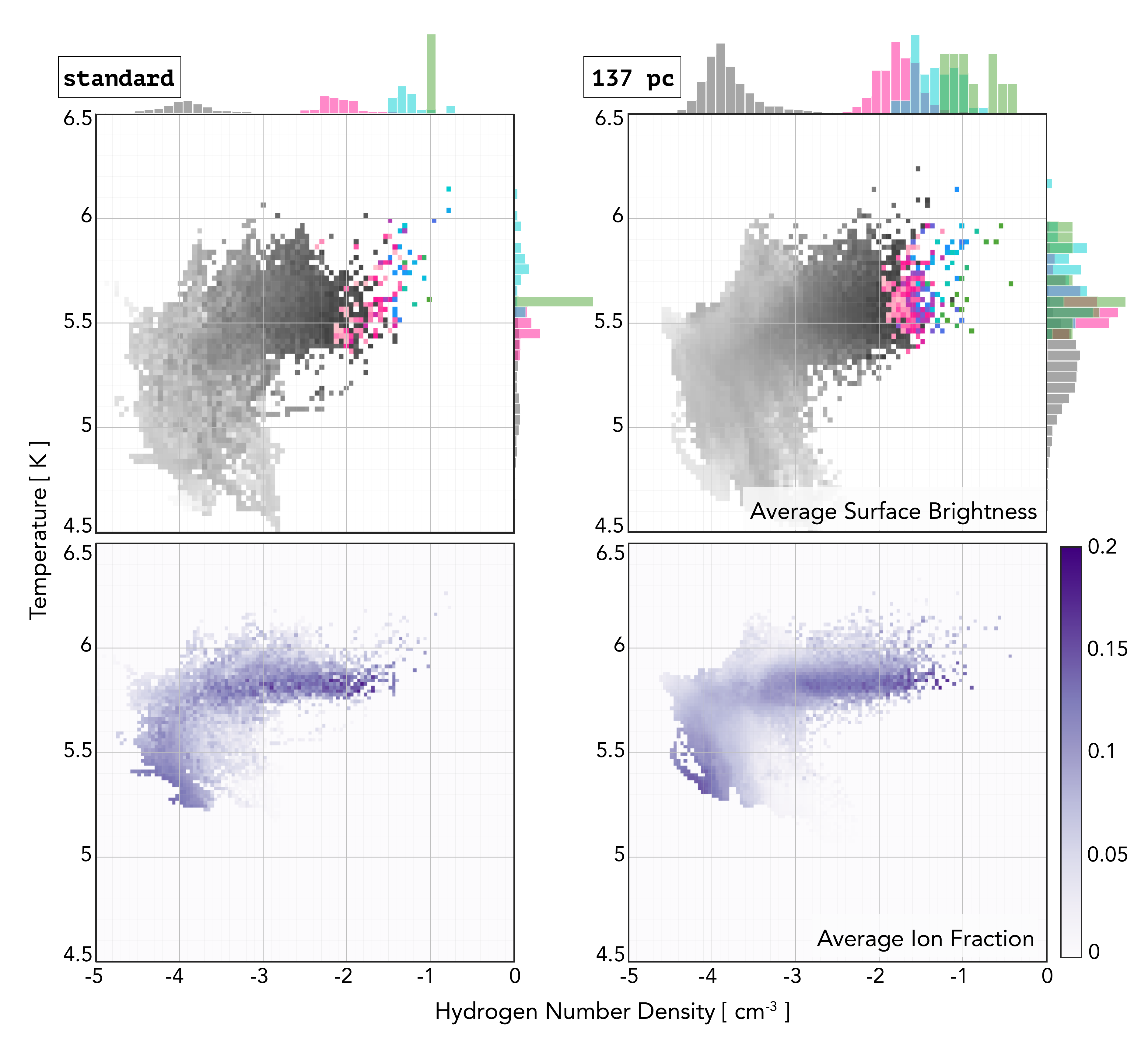}
\caption{Hydrogen number density ($n_{\mathrm{H}}$) and temperature, weighted by the \ovi\ emissivity. The left column shows the standard simulation and the right column the 137\,pc simulation. In the top panels, colors correspond to the average surface brightness of pixels in each bin. In the bottom panels, colors show the average \ovi\ ion fraction of pixels in each bin. The bottom panels show that high \ovi\ ion fractions are generated by both photoionization (low density, low temperature) and collionsional ionization (high density, high temperatures). However, only the collisional ionization, which occurs near the peak of the \ovi\ cooling curve, generates observable emission, as seen in the top panels.    \label{phase_OVI.fig}}
\end{figure}

It is a long-standing debate as to if the \ovi\ seen in absorption is predominantly photo- or collisionally-ionized \citep{tripp08,savage14,werk16,oppenheimer16,nelson18}. Figure \ref{phase_OVI.fig} show the hydrogen number density ($n_{\mathrm{H}}$) and temperature weighted by the \ovi\ emissivity along the line of sight for each pixel in the emission maps of Figure \ref{emis_map.fig}. In the top panels, the colors correspond to the average surface brightness of pixels that contribute to that bin, matching the color maps of Figures \ref{emis_map.fig}--\ref{sb_prof.fig}.  The normalized histograms show the distribution of $n_{\rm H}$ and temperature for pixels falling within a given detectability bin.  The phase diagrams show a clear trend that higher density leads to increasingly brighter emission. However, these dense regions also need to exist at the temperature at the peak of the cooling curve of that line to produce observable emission. Indeed, the observable bins all cluster around $T=10^{5.5}-10^6$ K for the \ovi\ line. 

Overall, there is not much variation between the two simulations in terms of the \ovi -emitting gas. The phase space is clearly more finely sampled by the higher resolution run, and a slightly wider range of densities and temperatures contribute to detectable pixels, most likely because the metallicity has increased for some of the pixels. 

The bottom panels show the same phase diagrams but colored to show the average ion fraction of pixels contributing to that bin. In both simulations, there is a large fraction of \ovi\ for hot, dense gas (top right of each panel) representing collisionally ionized gas. There is also a peak in the \ovi\ fraction at lower densities and at lower temperatures, revealing that there is also photoionzied \ovi\ gas in the simulation.  However, comparing the top and bottom panels, we can see that all of the gas that can be detected in emission comes from the collisionally ionized regime.

\subsection{The Effect of Angular Resolution on Deriving Physical Gas Properties}

\begin{figure}
\centering
\includegraphics[width=0.5\textwidth]{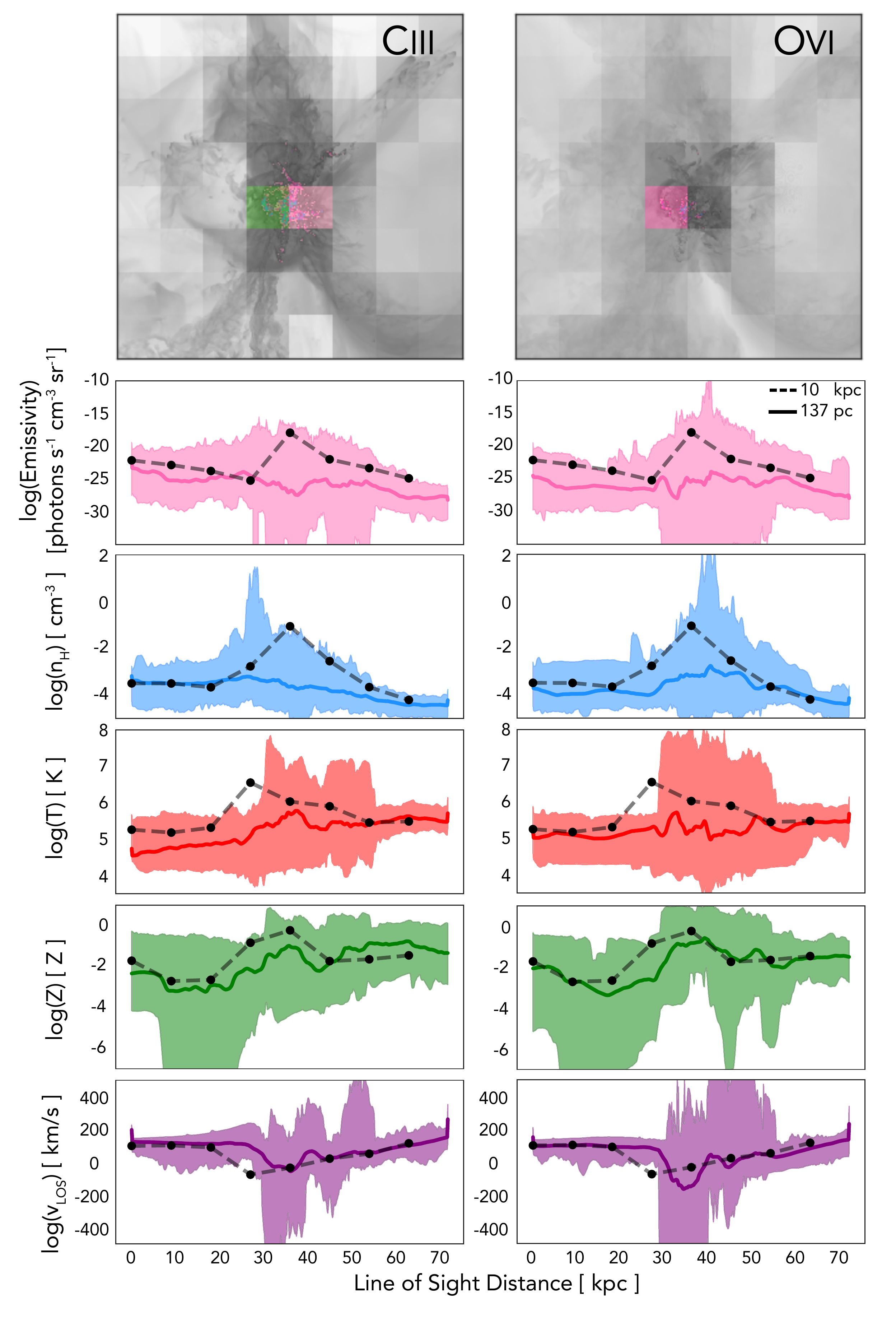}
\caption{Top panels show the emission maps of \ciii\ and \ovi\ for the 137\,pc simulation at its fiducial resolution and for an overplotted image where the resolution has been degraded to 10\,kpc. The pink pixel in both coarse images is found and the corresponding region in the high resolution image is identified. The LOS properties of the coarse simulation are then plotted in the lower panels as gray, dotted lines and of the highly refined simulation in solid colors. The solid colored line corresponds to the median values and the shaded region shows brackets the minimum and maximum value at each LOS position. The coarse resolution blends the gas physical properties such that the actual range of the gas's physical values, limiting what can be inferred from such a measurement.    \label{pixel_smear.fig}}
\end{figure}

Finally, the high resolution simulations can help place constraints on the degree to which the CGM properties are artificially blended by both coarse spatial resolution in the simulations and coarse angular resolution in the observations. The top panels of Figure \ref{pixel_smear.fig} show the emission maps for two lines, \ciii\ and \ovi, from the 137\,pc simulation and overplotted is the same image but where the pixel size is degraded to 10 kpc. The color map matches that of Figures \ref{emis_map.fig}--\ref{sb_prof.fig}. Visually, a single given observable pixel in the coarse image corresponds to a complex region with a large range of surface brightnesses and gas structures in the high resolution simulation. A single pixel wether simulated or observed is unable to capture such variations in CGM physical properties.

To understand this variation, we de-project the cube used to generate the emission map to recover the LOS information. We first identify the position where the pink pixel is found in the 10 kpc map and the corresponding region in the 137\,pc image. In the lower panels of Figure \ref{pixel_smear.fig}, we plot the physical properties along the LOS for the single pixel in the coarse map as gray, dashed lines. The line-of-sight variation of the emissivity, hydrogen number density, temperature, metallicity, and LOS velocity in the low resolution cube are evident.
For the set of pixels in the corresponding region of the full resolution cube, the colored lines show the median values of the physical properties along the LOS and the shaded regions correspond to the minimum and maximum values at each distance. The high resolution demonstrates that the coarser resolution in either simulations or observations blends the gas properties such that their variation is decreased. Gas is neither as hot or as cold, as dense or as diffuse, as metal-rich or metal-poor, as out-flowing or in-flowing in the coarse image as it is in the highly resolved image. 

Furthermore, the emission in a given 10\,kpc region is ultimately being driven by a handful of pixels that represent much smaller spatial scales. The brightest pixels can have emissivities of $10^{-15}$ to $10^{-10}$\,photons\,s$^{-1}$\,cm$^{-3}$\,sr$^{-1}$ as opposed to the median values of $10^{-25}$\,photons\,s$^{-1}$\,cm$^{-3}$ sr$^{-1}$.  How the properties of these bright pixels vary with the LOS and how these properties compare to what would be derived from {\sc cloudy} modeling of the measured emission on these scales will be the focus of future work.

\section{Instrument-specific Emission Maps} \label{sect:instr}

In this section, we re-present the surface brightness maps at $z=3$ of the 137\,pc simulation to reproduce the properties of two ground-breaking optical integral field units: KCWI on Keck and MUSE on the VLT. Direct detection of circumgalactic emission is one of the primary science goals for both of these instruments. Both have multiple observing modes, but we focus here on those which have the most sensitive surface brightness limits combined with the best angular resolution. This is the ``full-slice'' mode on KCWI and the ``narrow field'' mode on MUSE, the details of which we summarize in Table \ref{tab:instr}. 

\begin{table}

\centering
\begin{tabular}{ |c|c|c| } 
 \hline
\ & KCWI & MUSE \\ 
\hline
Mode Name & Full Slice & Narrow Field \\
FOV & 20\arcsec $\times$ 33\arcsec & 7.5\arcsec $\times$ 7.5\arcsec \\ 
Angular Resolution & 0.5\arcsec & 0.025\arcsec \\ 
Bandpass & 3500--5600 \AA & 4650--9300 \AA \\
Exposure Time & 30h  & 27h\\
SB Limit & $7\times 10^{-21}$ & $1\times 10^{-19}$ \\
\hline
\end{tabular} 
\caption{Summary of details of observing modes modeled in Section \ref{sect:instr} for KCWI and MUSE. Surface brightness limits are giving in ergs s$^{-1}$cm$^{-2}$arcsec$^{-2}$.} \label{tab:instr}
\end{table}

\begin{figure*}
\centering
\includegraphics[width=0.75\textwidth]{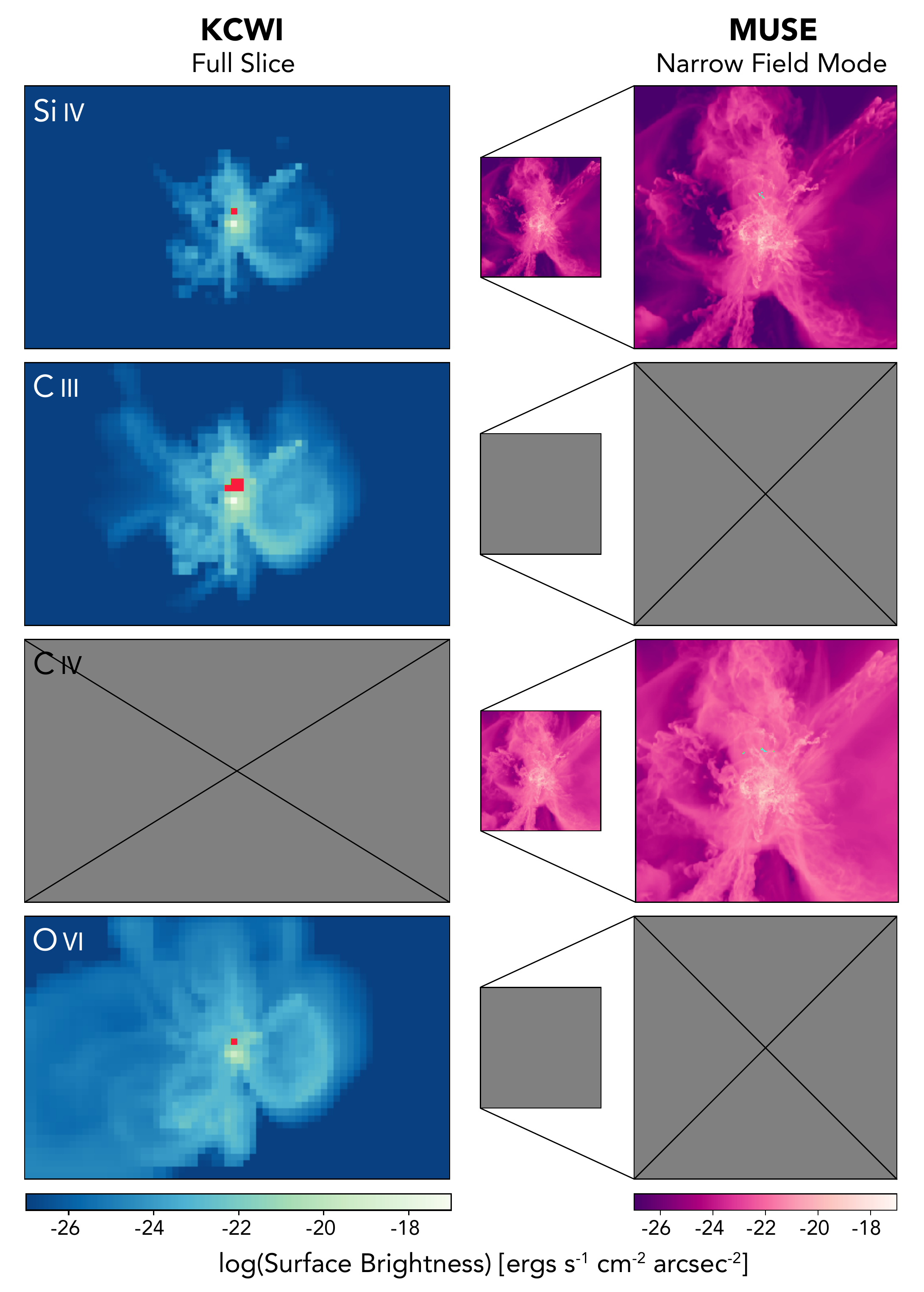}
\caption{Emission maps to match the properties of two specific observing modes on KCWI and MUSE as outlined in Table \ref{tab:instr}.  \emph{Pixels that lie above the surface brightness limit of the instrument are colored to stand out from the colormap}: red for KCWI and blue for MUSE.  Only a few pixels are detectable by either instrument. Gray boxes represent lines that have shifted out of the bandpass of the respective instrument and thus can not be observed at $z=3$. The large FOV of KCWI allows the entire CGM to be observed simultaneously. MUSE has a similarly broad observing mode but here we highlight the ``narrow field'' mode, which has exceptional angular resolution. Such high angular resolution allows for a detailed look at the gas properties that are only resolved in the simulation because of our new refinement scheme. \label{instr.fig}}
\end{figure*}

Figure \ref{instr.fig} shows the emission maps for the ions of interest at $z=3$ for both instruments. The relative sizes of the field of view (FOV) are depicted in the first two columns; the third shows a larger version of the MUSE images for clearer comparison with the KCWI images. All images reflect the stated angular resolution of the instruments' observing modes from their respective websites \footnote{https://www2.keck.hawaii.edu/inst/kcwi/configurations.html} \footnote{https://www.eso.org/sci/facilities/develop/instruments/muse.html}. For MUSE, the surface brightness limit is taken from \citet{wisotzki18} who observed in the wide field mode. In the narrow field mode we discuss here, the limits should be similar for all but readnoise-limited cases. However, we use this value as a good rule of thumb for this exposure time.  We focus on the narrow field mode since the small scales of the emission that are the focus of this work may raise the mean SB measured per spaxel as the emission is concentrated by the higher resolution of the instrument. 

The left panels of Figure \ref{instr.fig} show how the large FOV of KCWI in this mode (corresponding to $158\times 260$ physical kpc at $z=3$) allows the entire CGM be observed simultaneously.  In this way, a single observation can capture the processes shaping the inner and out CGM, whether that is cosmic filaments, minor mergers, or starburst-driven or AGN-driven outflows. 

MUSE has a mode the enables a FOV twice the size of the KCWI mode presented above, but here we have chosen to highlight the predicted performance of the instrument when operating with full adaptive optics. The superb angular resolution in the narrow mode allows for the details of the small-scale gas structure to be probed. The right panels of Figure \ref{instr.fig} demonstrate how both high spatial resolution in the simulations and high angular resolution in the observations is needed to understand the distribution of physical and spatial properties of the CGM as laid out in the previous sections. 

A major consideration that does not change with observing mode is the bandpass of the instruments. KCWI currently observes at much bluer wavelengths than MUSE. Because of the varying wavelengths of the emission lines, neither instrument can observe all of the metal lines presented here simultaneously. At lower redshifts, even more of the lines have shifted blue-ward of the MUSE bandpass. (H$\alpha$, which is detectable at $0 < z < 0.42$, is the notable exception.) 

Despite the FOV, bandpass, and angular resolution trade offs, both instruments are ultimately limited by their surface brightness sensitivities. For the panes in Figure~\ref{instr.fig}, pixels that are brighter than the limits of each instrument's observing mode are colored green. For both instruments and for any line, there are at most a handful of pixels that clear the detection limit. 

Binning (reducing angular resolution) or stacking (minimizing individual CGM features) may allow for better overall detection of the CGM emission. However, this single galaxy appears largely undetectable at $z=3$ for these instruments. In the next section, we discuss the implications of this for CGM emission studies overall. 

\section{Discussion} \label{sec:disc}
The instrument-specific emission maps shown above present a seemingly bleak picture for the future of directly detecting emission from the CGM. However, a more accurate statement is that they indicate that emission from \emph{this} galaxy remains out of reach.  While a Milky Way-like progenitor is interesting for understanding the evolution of galaxies like our own, this is not an ideal candidate to target for current emission observations.  This galaxy has a total mass of only $4\times 10^{10}$\,\Msun, has a star formation rate of 3--4\,\Msun\,yr$^{-1}$, and has no active AGN. A more massive galaxy will likely have a denser CGM, be fed by stronger cosmic filaments, and have more in-falling satellites to provide dense, stripped material throughout its halo. Higher star formation rates and AGN feedback will eject more mass and metals into the CGM as well as generate more radiation to enhance photoionization and can lead to strong time variability in the emission \citep{sravan16}. This effect is seen at low-redshift in the  COS-Bursts data low-redshift \citep{heckman17}. Thus, the prospects for more massive, active galaxies are promising for high-$z$ studies.

In addition to looking at galaxies with more observationally favorable properties, this work also looks towards the development of extremely large telescopes (ELTs) that may search for the CGM emission of progenitors of Milky Way-like galaxies. With larger collecting areas, ELTs can push to even lower SB limits with the same angular resolution as current large telescopes, increasing our chances of detecting galaxies such as the one presented in this paper. However, there will be trade-offs: if the typical solid angle of the sky sampled by these new instruments is significantly smaller (e.g., to take advantage of the extreme adaptive optics corrections on the ELTs), the sensitivity to diffuse gas may remain little changed. Studies such as this one can help evaluate such trade offs in future instrument designs in light of different science goals. 

Besides choosing galaxies with more favorable emission properties or lowering the surface brightness limit of observations, stacking remains a viable option for detecting emission from the CGM. While valuable information is lost pertaining to the exact gas distribution around each galaxy, stacking large numbers of galaxies shows that the extent of ionized gas is dependent on galaxy properties \citep{zhang18a} and can be used to probe the dominant source of ionization of the gas at different galaxy masses \citep{zhang18b}. Large-scale cosmological simulations could also be used to mimic such a stacking procedure and examine any biases due to viewing angles and time variability though that is beyond the scope of this paper. 

Furthermore, one of the biggest hindrances to detecting this emission is simply the distance and the resulting surface brightness dimming. Observing galaxies at lower redshift and in the UV, while still challenging, helps mitigate this particular limitation. \citet{corlies16} showed that emission from a Milky Way-like galaxy at $z=0$ can potentially be detected as far as 120\,kpc from the galaxy and that the covering fraction of detectable pixels can be as high as 5--10\%\ depending on the surface brightness limit assumed. Similar fractions are predicted for a larger, cosmological volume by \citet{bertone10}. UV-missions such as FIREBall-2 and LUVOIR may provide our most promising prospect for measuring the CGM in metal-line emission \citep{grange16,luvoir18}.

Finally, this paper has focused on metal-line emission because of its usefulness it tracing large-scale galactic gas flows and probing the ionization state of the CGM. Despite the limitations in interpreting its emission, \lyman\ is expected to be at least times brighter than the next brightest emission line \citep{bertone10}.  Future work will focus on combining these new, highly-refined simulations with a full radiative transfer code to make accurate predictions of \lya\ emission maps and kinematics. Similarly, although H$\alpha$ had the highest surface brightness, its long wavelength makes it unobservable by the optical IFUs we present here. However, this makes it a good candidate for observation with the {\em James Webb Space Telescope}; we will explore this potential in future work.

\section{Conclusions and Future Directions} \label{sec:conc}
Observing emission from the CGM would provide us with an unprecedented understanding of the 3D spatial and kinematic properties of how this gas is flowing into and out of galaxies, regulating their evolution. In this paper, we have focused on making metal-line emission predictions for the progenitor of a Milky Way-like galaxy at $z=3$. Our novel approach to resolving the CGM has allowed us to probe structures on scales smaller than ever before and to understand how the physical properties of these scales link back to observable gas. All of the results we present here owe to changes in the simulated circumgalactic resolution alone, with no changes to the resolution of the interstellar medium or sub-grid physics recipes.

Our main conclusions are:
\begin{enumerate}

\item High spatial resolution in the CGM is necessary to better predict its emission properties. Improved spatial resolution allows gas to clump on scales smaller than resolved by typical cosmological simulations. Many of these clumps are potentially detectable and found at larger distances from the galaxy than clumps in standard-resolution simulations. 

\item Globally, increasing the CGM resolution alone also increases the total luminosity of the lines considered here by an order of magnitude compared to the standard simulation. 

\item The differences in emission can be attributed to the broader range of physical properties the CGM possess once it is more finely resolved. More multiphase gas exists in the highly refined simulations at all distances from the galaxy as compared to the standard simulation. 

\item Two instrument-specific maps for observing modes on KCWI and MUSE show that the emission from a small, low star-forming, high-redshift galaxy is generally not detectable. Simulations like these can be used to identify better candidates for direct detection in the future.
\end{enumerate}

Moving forward, understanding the CGM will continue to be a science driver for future instrumentation, as it was for both KCWI and MUSE.  Interpreting new IFU observations that probe small angular scales requires more simulations like the ones we present here that can achieve small spatial resolutions in the halo. 

Future generations of FOGGIE simulations will include more massive galaxies as well as on those with more active merger and star formation histories. These systems will likely have a higher probability of detection of CGM emission from current instrumentation and provide a broader theoretical sample of highly-resolved galactic halos to guide target selection for future observations. 

Observing galaxies at lower redshift will also improve the likelihood of detecting this gas by decreasing the amount of SB dimming. Thus, future FOGGIE simulations will also focus on expanding the size of our refinement region to encompass the entire virial radius of galaxies at $z=0$ to make predictions for and inform the development of future UV observatories such as LUVOIR.

\acknowledgments
LC woud like to thank Britton Smith for helpful conversations throughout this project. We gratefully acknowledge the National Science Foundation for support of this work via grant AST-1517908, which helped support the contributions of LC, BWO, NL, JOM, and JCH. LC was additionally supported in part by HST AR \#15012.  BWO was supported in part by NSF grants PHY-1430152, AST-1514700, OAC-1835213, by NASA grants  NNX12AC98G, NNX15AP39G, and by HST AR \#14315. NL was also supported  by NASA ADAP grant NNX16AF52G. This work benefited from the successkid and sunglasses emoji on Slack. Computations described in this work were performed using the publicly-available Enzo code, which is the product of a collaborative effort of many independent scientists from numerous institutions around the world. Resources supporting this work were provided by the NASA High-End Computing (HEC) Program through the NASA Advanced Supercomputing (NAS) Division at Ames Research Center and were sponsored by NASA's Science Mission Directorate; we are grateful for the superb user-support provided by NAS. Resources were also provided by the Blue Waters sustained-petascale computing project, which is supported by the NSF (award number ACI 1238993 and ACI-1514580) and the state of Illinois. Blue Waters is a joint effort of the University of Illinois at Urbana-Champaign and its NCSA.

\facilities{NASA Pleiades, NCSA Blue Waters}

\software{astropy \citep{astropy2},  
          {\sc cloudy} \citep{ferland13}, 
          Enzo \citep{bryan14},
          grackle \citep{smith17},
          yt \citep{turk11}
          }

\bibliographystyle{aasjournal}
\bibliography{references}

\end{document}